\documentclass{emulateapj}

\newcommand{\iso}{{\em ISO}}
\newcommand{\iras}{{\em IRAS}}
\newcommand{\mum}{\ifmmode{\rm \mu m}\else{$\mu$m}\fi}

\submitted{Submitted to the Astrophysical Journal, submitted 18 Dec., 2004; 
revised 22 Jul., 2005; accepted 14 Sep., 2005}
\journalinfo{ }

\begin{document}

\title{The unusual silicate dust around HV 2310, 
an evolved star in the LMC}

\author{
G.~C.~Sloan\altaffilmark{1}, 
D.~Devost\altaffilmark{1},
J.~Bernard-Salas\altaffilmark{1},
P.~R.~Wood\altaffilmark{2},
J.~R.~Houck\altaffilmark{1} 
}
\altaffiltext{1}{Cornell University, Astronomy Department,
  Ithaca, NY 14853-6801, sloan@isc.astro.cornell.edu}
\altaffiltext{2}{Research School of Astronomy and Astrophysics, 
  Australian National University, Cotter Road, Weston Creek ACT 2611, 
  Australia, wood@mso.anu.edu.au}

\begin{abstract}

The spectrum of HV 2310, an evolved star in the Large 
Magellanic Cloud, taken with the Infrared Spectrograph 
(IRS) on the Spitzer Space Telescope reveals the 
presence of an optically thin shell of silicate dust with
unusual spectral structure in the 10~\mum\ feature, with an
emission peak at 9.7~\mum, a saddle at 10.4~\mum, and an 
extended shoulder to 11.2~\mum.  This structure is similar to
spectra from crystalline silicate grains, and of the available
optical constants, forsterite provides the best fit.
The spectrum also shows structure at 14~\mum\ which may 
arise from an unidentified dust feature.

\end{abstract} 


\section{Introduction} 

The advances in sensitivity made by the Infrared 
Spectrograph\footnote{The IRS was a collaborative venture 
between Cornell University and Ball Aerospace Corporation 
funded by NASA through the Jet Propulsion Laboratory and the 
Ames Research Center.} \citep{hou04} on the {\it Spitzer} 
Space Telescope \citep{wer04} expands the field of infrared 
spectroscopy to distant, faint, metal-poor galaxies.  The 
absorption and emission of radiation by dust in these galaxies 
dominates the astrophysics in the infrared, and if we are to 
understand the nature of this dust, we must determine if global 
dust properties depend on metallicity, and if so, how.

The Magellanic Clouds provide the ideal laboratory for this 
investigation, as the global metallicity of the Large 
Magellanic Cloud (LMC) is approximately 0.5 times solar, and 
in the Small Magellanic Cloud, the metallicity is 0.25 times
solar \citep[e.g.][]{hi97, lu98}.  In the Milky Way, stars 
on the Asymptotic Giant Branch (AGB) account for the majority 
of the dust injected into the interstellar medium 
\citep[ISM;][]{geh89}, so it is reasonable to assume a similar 
case in the Magellanic Clouds.  The dust in shells around 
evolved stars has not yet been mixed into the interstellar 
medium, so the study of spectra from circumstellar dust shells
enables the different chemical components of the future ISM 
to be identified and studied separately.

Here we report on one of the more interesting sources in our
study of the Magellanic Clouds, HV 2310, a Mira variable in 
the LMC with a 598-day period.  We show that the silicate 
emission from HV 2310 is quite unusual, and that it the
presence of crystalline silicates, particularly forsterite,
can explain this structure.

\section{Observations and Reductions} 

Spitzer observed HV 2310 on 2004 April 18 (IRS Campaign 5) 
with the Short-Low (SL) and Long-Low (LL) modules.  The  
total integration times were 112 seconds for SL and 480 
seconds for LL.  The analysis begins with the flatfielded 
images generated by the standard online data-reduction 
pipeline at the {\it Spitzer} Science Center (S11.0) and 
follows the procedure described by \cite{sl04}.  The K 
giant HR 6348 served as the standard for SL, while it and 
the K giants HD 166780 and HD 173511 calibrated LL.  The 
wavelengths from the S11.0 pipeline have been shifted 
slightly to account for the slight offsets described by 
\cite{sl05}.  In SL order 1, the offset is 0.042~\mum.

\section{Analysis} 

Figure 1 presents the spectrum of HV 2310, which includes
contributions from the star and the silicate dust in its 
circumstellar shell.  \cite{es01} found that dust shells
which produce similar spectra in the galactic sample are
optically thin.  We assume the optically thin case for HV
2310, which means the contribution from the star and dust 
can be separated using the method developed by \citet[][ 1998; 
hereafter SP]{sp95} to spectra from the Low-Resolution
Spectrometer on \iras.  They fit a stellar continuum to 
the spectrum in the 7.64--8.70~\mum\ range and subtract it 
from the entire spectrum.  The difference is the dust 
spectrum.

For the galactic sample of AGB stars examined by SP,
NU Pav served as a proxy for the stellar component in the 
spectrum from each source.  NU Pav is an M6 giant, 
and its spectrum can be approximated with a 3240 K Engelke 
function \citep{en92} and a SiO fundamental band with an 
absorption depth of 15\% (measured at low resolution).  The 
spectrum of NU Pav used here was observed by the 
Short-Wavelength Spectrometer (SWS) on \iso\ (TDT 12103028).  
We used the spectrum from the SWS Atlas \citep{sl03a}, 
binned to the IRS resolution.  We modified the wavelength
range for fitting the stellar continuum by SP to 5.5--7.0~\mum.

\begin{figure} 
\includegraphics[width=3.5in]{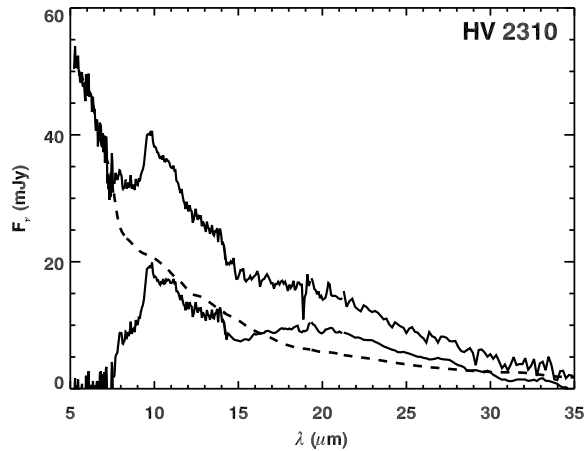}
\caption{The IRS spectrum of HV 2310 (upper solid line) with
a fitted continuum from NU Pav (dashed line) and the residual 
dust spectrum (smoothed past 14~\mum; lower solid line).  The 
apparent feature at 18.8~\mum\ is an artifact.}
\end{figure}

The only apparent features in the stellar-subtracted spectrum 
are solid-state in nature, so when plotting the 
continuum-subtracted spectrum in Figure 1, we smoothed the LL
data using a seven-pixel boxcar.  Subsequent figures plot the 
spectra at all wavelengths with no smoothing.

A silicate emission feature at 10~\mum\ dominates the dust 
spectrum of HV 2310, and it has a double-peaked structure,
with an emission peak at 9.7~\mum, a minimum at 10.4~\mum,
and an extended shoulder to 11.2~\mum.  This basic structure
appears in both nod positions.  Similar spectral shapes 
appear in dust spectra from comets and young stars and are 
attributed to emission from crystalline silicates 
\citep[e.g.][]{bre87, kna93, hon03}, but such a shape is 
rarely seen in spectra from dust shells around evolved stars.  

\subsection{Comparison to the galactic sample} 

SP classify optically thin silicate emission (SE) features
by using the ratios of the dust emission at 10, 11, and 
12~\mum\ (after removing the stellar contribution).
Roughly, the SE classification $= 10 F_{11}/F_{12} - 7.5$.
The classifications run from SE1, which arises from
amorphous alumina grains, to SE8, the classic silicate
feature at 10~\mum\ produced by amorphous silicates.  

Following the same method, we classify the dust spectrum 
of HV 2310 as SE6, although the spectrum is close to the 
boundary with SE5.  SE6 is at the transition to structured 
silicate emission spectra (SE4--6), which show broader 
silicate features, often with a shoulder at 11~\mum\ and in 
some cases, with a strong 13~\mum\ feature (suffix ``t'').  
Subtracting an Engelke function without SiO absorption 
moves the classification over the boundary to SE5.  This 
would be the extreme metal-poor case, and it has little 
other effect on our analysis.

\begin{figure} 
\includegraphics[width=3.5in]{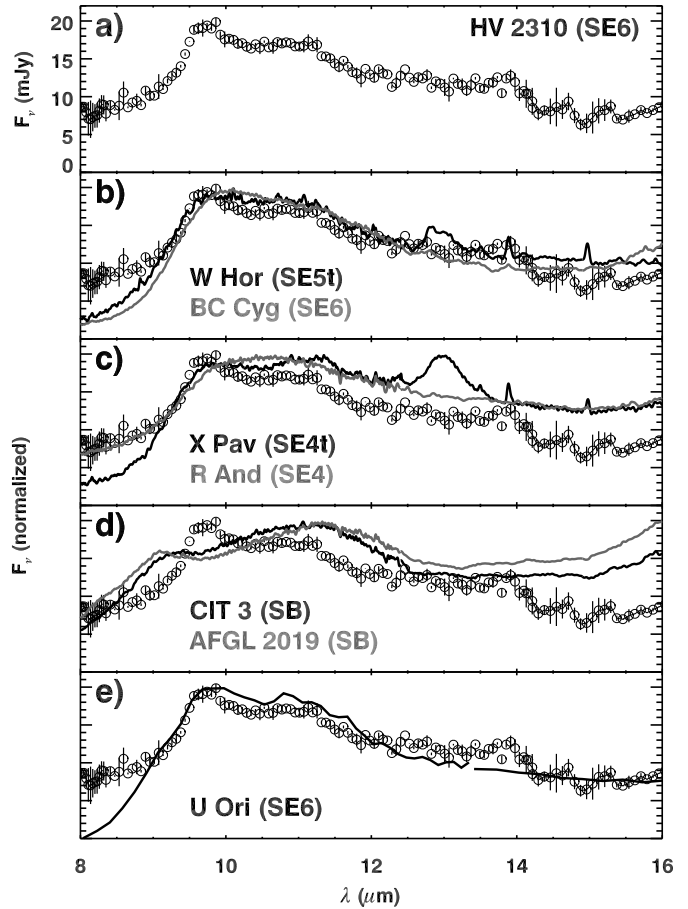}
\caption{The smoothed dust spectrum of HV 2310 (Panel {\it 
a}) compared to continuum-subtracted spectra from the SWS 
on \iso\ (Panels {\it b-d}) and the LRS on \iras\ (Panel {\it 
e}).  Panel {\it b} compares HV 2310 ({\it circles}) to 
spectra from the SWS in nearly the same position on the 
silicate dust sequence, both with and without 13~\mum\ 
emission features.  Panel {\it c} makes a similar comparison 
to SE4 spectra, which tend to show more structure.  
Panel {\it e} compares HV 2310 to U Ori, the one source 
in the LRS sample with an SP classification and similar
flux ratios from the dust at 9.7, 10.3, and 10.7~\mum\ 
which are similar to those from HV 2310.}
\end{figure}

Figure 2 compares the dust spectrum of HV 2310 with similarly
extracted dust spectra from evolved stars in the Galaxy.
Panel {\it b} compares HV 2310 with two sources with similar
flux ratios at the boundary between SE5 and SE6, both with and
without a 13~\mum\ feature.  The spectrum of W Hor (the
13~\mum\ source) resembles HV 2310 more closely, with a clearer
shoulder at 11.2~\mum\ than BC Cyg.  Panel {\it c} compares
HV 2310 with two SE4 sources, which show a larger contribution
to the dust emission at 11~\mum.  The 13~\mum\ source, X Pav,
shows a clear minimum at 10.5~\mum\ and an extended shoulder
out to 11.4~\mum, while R And differs more from HV
2310.  The two 13~\mum\ sources in Figure 2 show some 
similarities with HV 2310, but neither has the narrow 9.7~\mum\
feature or as strong of a dip from 9.7 to 10.4~\mum.  Among
these SWS spectra, which were chosen because of their general
similarities to HV 2310, HV 2310 is unique.

There are competing explanations of how the structured silicate
spectra in the SE4--6 range are produced.  Many authors have 
proposed crystalline silicate grains 
\citep[e.g.][]{tie90,lml90,nh90}, but other explanations are
also possible.  \cite{lmp00} fit models mixing amorphous alumina 
and amorphous silicate grains to the structured silicate spectra.
While such a combination might reproduce the spectrum of R And
in Figure 2, it could not produce the structure in the spectrum
of X Pav.  \cite{es01} showed that self-absorption of the 
10~\mum\ silicate feature would also work.  The dust shells 
producing structured silicate spectra would have to be optically
thick to self absorb, but then would have to be geometrically
thin to match the observed [12]$-$[25] colors.

Panel {\it d} in Figure 2 compares the spectra of two 
self-absorbed amorphous silicate features (classification SB) 
to HV 2310.  \cite{es01} show that as the silicate feature goes 
into self-absorption, the silicate feature broadens, and then 
the 10~\mum\ emission drops while the 11~\mum\ emission remains 
strong.  As the optical depth increases, the spectrum develops
a secondary peak in the vicinity of 9~\mum.  In self-absorbed
silicate spectra, the short-wavelength wing of the 10~\mum\
feature is always weaker than the long-wavelength wing, but in
HV 2310, it is the other way around.

The flux ratios used to classify HV 2310 as an SE6 do not 
measure the dip in emission between 10 and 11~\mum.
Shifting the wavelength regions used to determine the flux
ratio so that they are centered at 9.7, 10.3, and 10.7~\mum\ 
allows us to compare HV 2310 to the larger sample of LRS
spectra from evolved stars in the galaxy examined by SP.
We limited our comparison to the $\sim$ 200 sources with 
[12] $ < -1$ and a dust contrast of 8\% or more with respect 
to the continuum from 7.67 to 14.03~\mum.  U Ori is the one 
source in the LRS database with similar flux ratios in these
modified wavelength regions and a similar SP classification.
The comparison in Panel {\it e} shows that the spectra have
much in common, but in U Ori, the peak at 9.7~\mum\ and 
shoulder at 11.2~\mum\ are not as well pronounced.  

\subsection{Possible amorphous dust components} 

\citet[][ hereafter OHM]{ohm92} generated optical constants for 
amorphous silicate grains which reproduce the spectra observed
in typical shells around evolved stars.  Two other amorphous 
grains commonly identified in oxygen-rich circumstellar 
environments are amorphous alumina \citep{var86,ona89} and 
glassy pyroxenes \citep{gh86,jag94}, although the latter have 
only been noted around young stars, not evolved stars.  
Adding these optical constants to the OHM constants will
extend the peak of the 10~\mum\ feature to the blue and add
a shoulder to the red, which should force the shape of the 
10~\mum\ silicate feature to more closely resemble the 
observed feature in HV 2310.

\begin{figure} 
\includegraphics[width=3.5in]{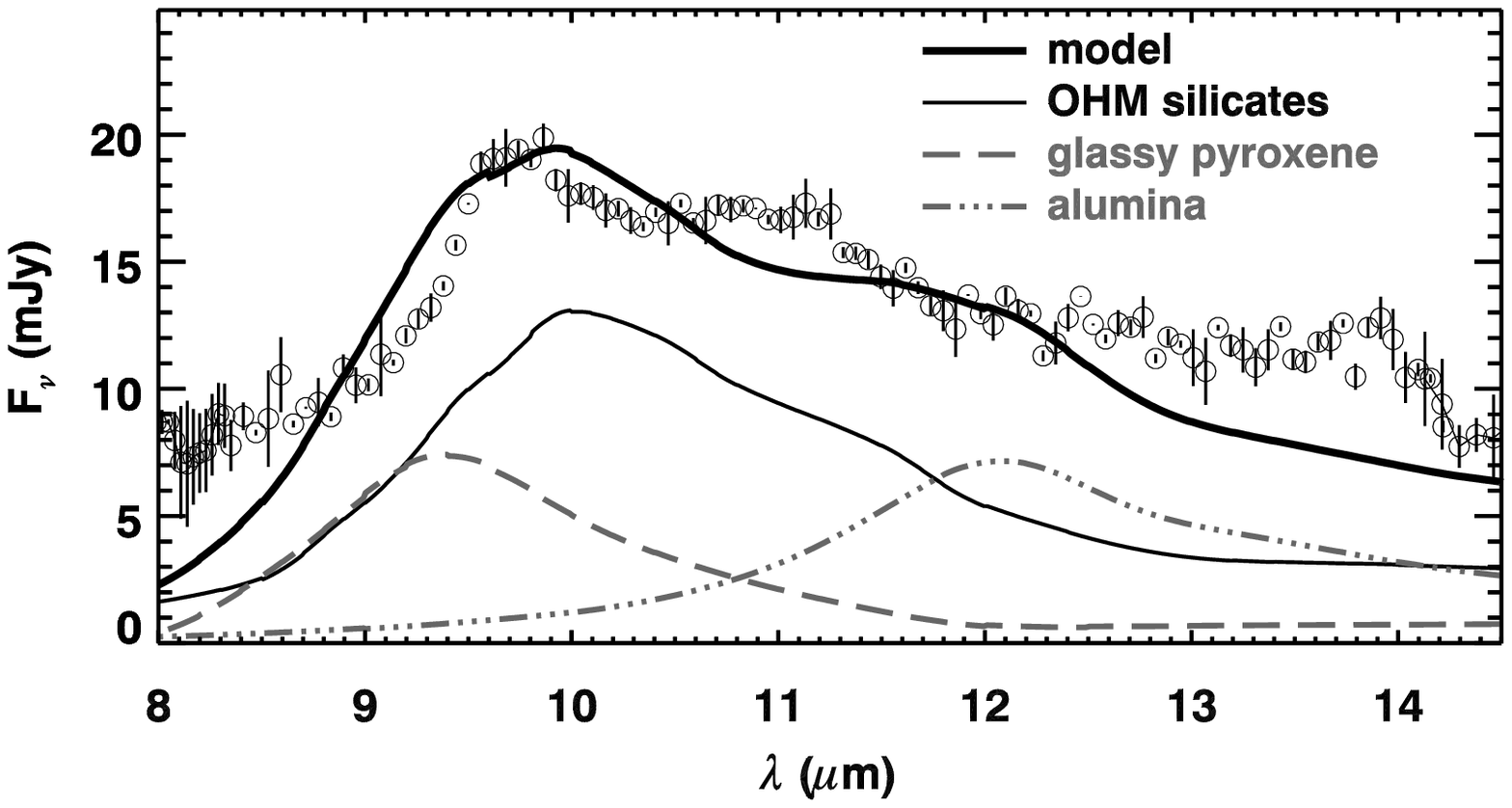}
\caption{A model using amorphous grain species ({\it 
solid, dashed, and dotted curves}) fit to the dust spectrum
of HV 2310 ({\it circles}).  The model is calculated 
assuming 450 K grains based on the optical constants of 
\citet[][ Set 1, 70\%]{ohm92}, 
\citet[][ 95\% Mg, 20\%]{jag03}, and
\citet[][ ISAS sample, 10\%]{ko95}.
The glassy pyroxene feature is too broad and too blue
to fit the observed spectral feature at 9.7~\mum, and
amorphous alumina peaks too far to the red of the 
11.2~\mum\ shoulder.}
\end{figure}

Figure 3 shows a model of the spectrum of HV 2310 using 
70\% OHM silicates, 20\% glassy pyroxene, and 10\% amorphous 
alumina.  To model the dust emission, we generate optical 
efficiencies ($Q$) from the complex indices of refraction 
following the technique described by \cite{dl84}.  For 
amorphous silicates, we use OHM Set 1.  The ISAS sample of 
amorphous alumina measured by \cite{ko95} match the broad 
features (SE1--3) observed in the galactic sample well 
\citep{es01}.  For glassy pyroxenes, we adopt the 95\% Mg 
sample measured by \cite{jag03}.  Shifting to lower 
percentages of Mg broadens the emission feautre at 9.4~\mum, 
but it does not shift its position significantly.  We assume 
a grain size of 0.1~\mum\ and a grain temperature of 450 K.  

The pyroxene adds more flux to the blue of the OHM peak at 
10.0~\mum, but the pyroxene peaks at 9.4~\mum\ and has a full 
width at half maximum (FWHM) of 1.6~\mum, compared to the 
actual feature in HV 2310, which peaks at 9.7~\mum, and is 
narrower than the model pyroxene.  By fitting a linear 
continuum under the feature in HV 2310 from 9.0 to 11.3~\mum, 
we measure the FWHM to be 0.7~\mum.\footnote{Fitting two 
gaussians to the total profile centered at 9.6 and 11.0~\mum\ 
gives a FWHM for the 9.6~\mum\ component of 1.1~\mum, but this 
is still narrow compared to the glassy pyroxene, and it is 
too simplistic, as it ignores the presence of the 10~\mum\ 
silicate feature.}  Adding more pyroxene would only increase 
the mismatch between the model and the dust spectrum from 
HV 2310 between 9.0 and 9.4~\mum.

The alumina peaks at 12.0~\mum, too far to the red to
reproduce the observed shoulder extending to 11.2~\mum.  
Thus, we conclude that glassy pyroxenes and amorphous alumina 
cannot reproduce the observed spectral structure produced by 
the dust around HV 2310 between 9.4 and 11.2~\mum.  However, 
as shown below, these dust species may still be present in 
the circumstellar shell of HV 2310.

\subsection{Crystalline silicates} 

The spectral structure in the dust emission from HV 2310
cannot result from a combination of amorphous alumina and
amorphous silicates, and it cannot arise from self-absorbed
emission from amorphous silicate grains.  The dust must
contain some other component, and given the narrowness of
the emission feature at 9.7~\mum, this component is most
likely crystalline.  To test this possibility, we will construct 
a simple model which is a sum of optically thin emission from 
amorphous dust grains and crystalline silicate grains.  

Crystalline silicate grains show a wide variety of spectral
structure depending on the olivine-to-pyroxene ratio, the
Mg/Fe ratio, and the shape of the grains (among other 
variables such as size and fluffiness).  Two laboratories
(in Jena and Kyoto) have produced particularly useful sets of 
data on silicates in the olivine and pyroxene series.  

The Kyoto group have published a series of opacity 
measurements for the olivine series, covering several Mg/Fe 
ratios from forsterite (the Mg-rich endmember, Mg$_2$SiO$_4$) 
to fayalite (the Fe-rich endmember, Fe$_2$SiO$_4$) 
\citep{ko03}.  They have also published a similar series of 
measurements for pyroxenes from enstatite (MgSiO$_3$) to 
ferrosilite (FeSiO$_3$) \citep{chi02}.  

\begin{figure} 
\includegraphics[width=3.5in]{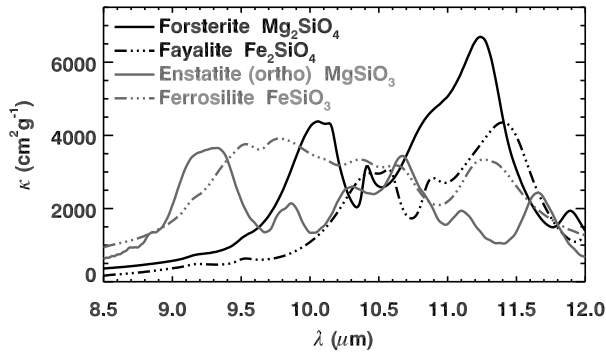}
\caption{Opacities for the endmembers of the olivine series
(forsterite and fayalite) and the pyroxene series (enstatite
and ferrosilite) from \cite{ko03} and \cite{chi02},
respectively.  Only forsterite is consistent with the 
dust spectrum of HV 2310, although the emission peaks in the
laboratory data are to the red of the observed peaks.}
\end{figure}

Figure 4 presents the opacities from the Kyoto group for the 
endmembers of both series.   Forsterite, with its peaks at 10.0
and 11.2~\mum, fits the dust spectrum from HV 2310 better than
the other dust species.  Fayalite produces features at 10.5 and 
11.4~\mum, while ferrosilite has a very broad feature centered
at 9.7~\mum\ and a secondary peak at 11.3~\mum.  Both these
dust spectra would fill in the 10.4~\mum\ saddle in HV 2310 
and can thus be ruled out.  Enstatite produces features at 9.2, 
10.7, and 11.7~\mum.  The strongest feature is at 9.2~\mum\ 
and is not observed in HV 2310.

Forsterite is not an ideal match to HV 2310, as the 10.0~\mum\ 
feature is to the red of the observed 9.7~\mum\ feature, and 
the 11.2~\mum\ peak can account for only the red portion of
the observed shoulder from 10.4 to 11.2~\mum.  However, the 
wavelengths of the spectral features from crystalline dust 
species depend strongly on grain shape.  While the opacity 
measurements from the Kyoto group provide good sampling over 
several Mg percentages from 100\% to 0\% for both olivines and 
pyroxenes, they do not measure the complex indices of 
refraction or dielectric constants.  As a consequence, we 
cannot use them to analyze the effect of grain shape on the 
dust spectra.

Fortunately, the Jena group provides indices of refraction
for forsterite and fayalite \citep{fab01}, as well as 
enstatite \citep{jag98}.  They also provided opacities
(mass absorption coefficients) for spherical grains, a
continuous distribution of ellipsoids (CDE1), and a 
modified CDE using quadratic weighting to emphasize spherical 
grains (CDE2).  Generally, CDEs are unsatisfactory because
they include grains which are infinitely cylindrical or flat.
On the other hand, deviations from spherical symmetry are
essential to reproduce observed interstellar polarizations 
\citep[e.g.][]{km95}.  To generate optical efficiencies for
a variety of grain shapes, we apply the methodology presented 
by \cite{fab01}. 

\begin{figure} 
\includegraphics[width=3.5in]{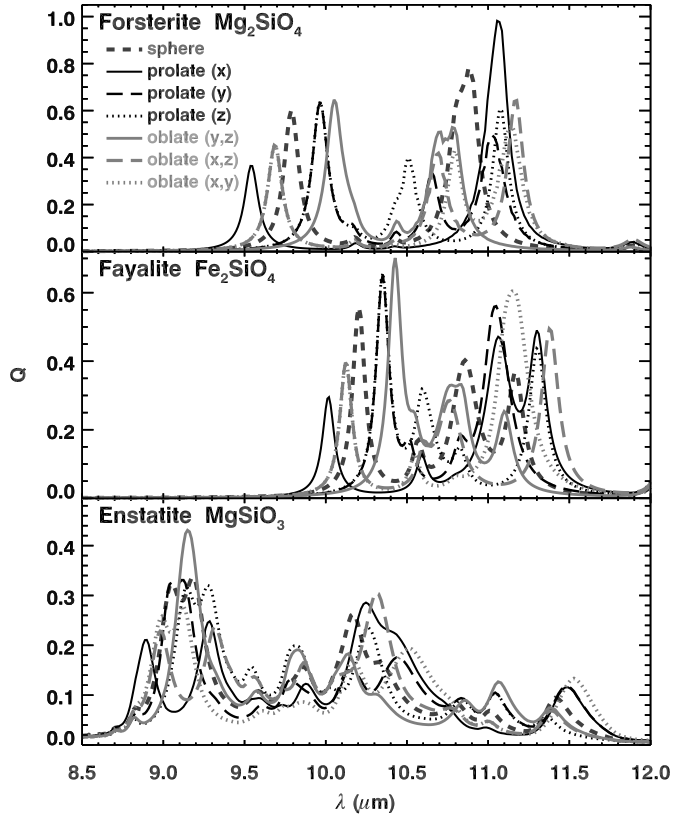}
\caption{Optical efficiencies for spheroids from forsterite,
fayalite, and enstatite.  The spheroids have 3:1 axis ratios
with the long axes as indicated in the legend of the top 
panel.  The optical efficiences are based on complex indices 
of refraction by \citet[][ top two panels]{fab01} and
\citet[][ bottom panel]{jag98}.  Fayalite and enstatite cannot
reproduce the structure in the dust spectrum of HV 2310 from
9.7 to 11.2~\mum.  With the right shape distribution, it is
possible to fit the spectral structure with forsterite.}
\end{figure}

Figure 5 explores the effect of changing the grain shape 
among all possible spheroids having 3:1 axial ratios.  The
lengths along the axes are normalized such that the volume
equals that for a sphere of radius 0.1~\mum.  Panel (a)
shows that varying the grain shape can move the feature
seen at 9.8~\mum\ in spherical grains from 9.5 to 10.1~\mum.
Similarly, the 10.9~\mum\ feature can shift from 10.5 to 
11.2~\mum.  Thus, with the right choice of shape 
distribution, it should be possible to fit a model 
including forsterite to the observed dust spectrum of 
HV 2310.  Panels (b) and (c) effectively rule out fayalite
and enstatite, as neither can produce a featur at 9.7~\mum\
as observed in HV 2310.  Increasing the axial ratios to
10:1 does not improve matters.

\subsection{Fitting crystalline silicates to HV 2310} 

\begin{figure} 
\includegraphics[width=3.5in]{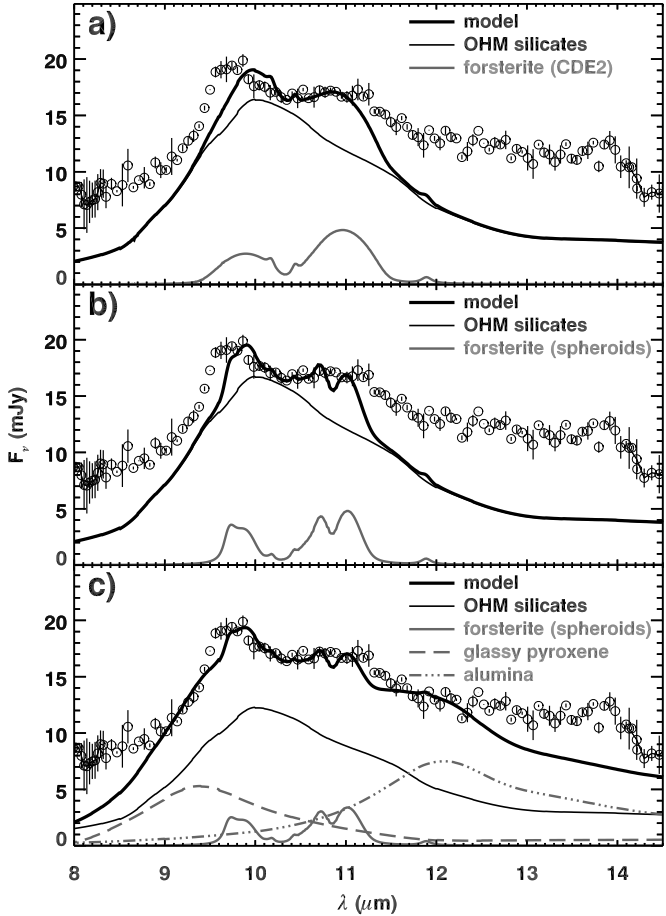}
\caption{The dust spectrum of HV 2310 ({\it circles}), 
compared to optically thin dust models using OHM silicates 
and other dust species ({\it lines}).  Panel (a) includes
a CDE2 distribution of forsterite (8\%) and OHM silicates
(92\%).  Panel (b) changes the shape distribution
of forsterite so that it now runs from prolate (z) to
spherical to oblate (x,z), weighted such that 90\% of the 
grains have axial ratios less than 3:1.  This distribution
requires 7\% forsterite and 93\% OHM silicates, and while
it improves the fit between 9.8 and 11.0~\mum, it still
does not account for the wings of emission to either side.
Panel (c) adds glassy pyroxene (15\%) and amorphous
alumina (11\%) to the forsterite (5\%) and OHM silicates
(69\%) and fills in most of the wings well.  The 
possible emission feature at 14~\mum\ is now readily 
apparent.}
\end{figure}

The top panel in Figure 6 compares a model combining OHM 
silicates (92\%) with a CDE2 distribution of forsterite (8\%)
to the dust spectrum of HV 2310.  Qualitatively, the
model reproduces both the narrow short-wavelength peak
and the broader long-wavelength shoulder, but the fit
is far from perfect quantitatively.  The model produces
a peak which is to the red of the observed 9.7~\mum\ 
peak, and it does not produce a shoulder which continues
to 11.2~\mum.

The top panel in Figure 5 shows that extending grains 
along the x axis shifts the 9.8~\mum\ feature to the blue,
while shortening the x axis shifts the feature to the red.
The y and z axes behave similarly, but in the opposite
direction.  By simply varying the size of the grains along
any one axis with respect to the other two, we can broaden
the 9.8~\mum\ feature without shifting it to better match 
the observed profile.  The behavior of the 10.9~\mum\
feature is more complicated.  Deviations from spherical 
symmetry split this feature into two components, which
should reproduce the observed plateau from 10.5 to 11.2~\mum.
Only oblate grains extended along the x and z axes are able 
to push the red component to 11.2~\mum.  Similarly, only
prolate grains extended along the z axis can push the blue
component to 10.5~\mum.

Thus, a distribution of spheroids including prolate (z) 
grains and oblate (x,z) grains should improve the fit 
provided by the CDE2 model by building a better plateau
and broadening the 9.8~\mum\ feature without shifting it
to longer wavelengths.  To test this possibility, we 
construct a spheroidal distribution running from prolate (z) 
through spherical to oblate (x,z), using gaussian weighting 
such that 90\% of the grains have axial ratios less than 3:1.

The purpose of this exercise is {\it not} to determine
the actual shape distribution of the grains around HV 2310,
but simply to demonstrate that changing the shape 
distribution can improve how well forsterite fits the
observed spectral structure.  As Panel (b) of Figure 6 shows, 
the chosen spheroidal distribution improves the model in the 
vicinity of the 9.7~\mum\ feature (although there is still
room for further improvement) and it fits the plateau to 
11.2~\mum\ about as well as the CDE2 model.  In this model, 
7\% of the grains are crystalline.

In Panels (a) and (b) of Figure 6, the models underestimate
the observed flux in the wings to either side of the 
structure from 9.5 to 11.2~\mum.  Panel (c) shows that adding 
the glassy pyroxene and amorphous alumina samples first 
considered in \S 3.2 improves the model substanitally, 
suggesting that these dust species may be present in the 
shell around HV 2310.  

It is important to emphasize the limitations of the
modelling illustrated in Figure 6.  The relative spectral
positions of the dust features depend on the shape of the
grains, but they also depend on the relative amounts of 
Mg and Fe.  As Figure 5 shows, fayalite produces similar
spectra to forsterite, but each feature is shifted to the
red, with the 9.8~\mum\ feature shifting more than the 
10.9~\mum\ feature.  While \cite{ko03} measured samples 
intermediate between forsterite and fayalite (at 
$\sim$20\% intervals), they only provide opacities.  
\cite{fab01} provide indices of refraction, but only for
the endmembers of the olivine series.  Thus, we are not
in a position to fit both the Mg/Fe ratio and the shape
distribution simultaneously.  It is unlikely that the
dust in HV 2310 has no Fe content, making it desirable
to have optical constants for Mg-dominated but not Mg-pure
crystalline olivine species.  Without these constants, it
is unreasonable to further contrain the shape distribution
of the grains.  To repeat, the objective here is to
demonstrate the feasibility of fitting dust species near
the forsterite end of the olivine series to the dust
spectrum of HV 2310.

\subsection{Other spectral features} 

The spectrum of HV 2310 shows other spectral features as
well.  It has a silicate feature at 18--20~\mum\ which 
peaks at $\sim$19~\mum.  The precise center is difficult 
to pin down due to the noise apparent in the unsmoothed 
spectrum.  The spectrum of HV 2310 shows sharp inflections 
at 13.9~\mum\ and 14.3~\mum, which can be interpreted 
either as a dust emission feature at 13.9~\mum\ or an 
absorption band with its blue wing starting at 13.9~\mum.  
CO$_{2}$ gas should absorb in this spectral region.  Models 
by \citet[ Appendix A]{cam02} show that if the temperature 
of the gas is low ($\sim$500 K), the band begins at 
$\sim$13.5~\mum, but the models do not show a sharp 
inflection as seen in our spectrum, and the models show a 
strong contribution from the $\nu_2$ bending mode at 
14.98~\mum\ which does not appear in our spectrum.  

In Panel (e) of Figure 2 and Panel (c) of Figure 6, the
spectral structure at 14~\mum\ appears to be above any
possible dust continuum, leading us to suspect that we
have observed a new dust feature from an unknown carrier.
Our confidence is limited by the coincidence of this
feature with the boundary between SL and LL.  One inflection
in the slope appears in each module, and where they overlap,
the slope is in good agreement, but we await confirmation of
this feature in other sources and with the Short-High module
of the IRS.

The spectrum of HV 2310 does not show a 13~\mum\ emission
feature, nor does it show the dust emission features at 20 and
28~\mum\ or the CO$_2$ bands in the 13--16~\mum\ range, which 
\cite{sl03b} found to correlate with the 13~\mum\ feature in
their study of optically thin dust shells around evolved stars
in the Galaxy.  The dust spectrum does show gentle inflections 
at 23, 28, and 33~\mum, and while these are at wavelengths where 
crystalline silicate emission would be expected, the spectrum 
also shows an inflection at 30~\mum, where nothing would be 
expected.  None of these features are clear in the unsmoothed 
spectrum, making us cautious about their significance.

The absence of the 13~\mum\ feature in the spectrum of
HV 2310 is interesting. \cite{sl03b} have suggested that
the leading candidate for the carrier of this feature is
crystalline alumina.  We see crystalline silicate features
in the 10--11~\mum\ region, and amorphous alumina could
well be present in the 12~\mum\ region, so why is 
crystalline alumina absent at 13~\mum?  The answer to
this question is not clear, but it may lie in the possible
dust feature at 14~\mum.

\section{Discussion} 

The dust emission in the spectrum of HV 2310, with its narrow
9.7~\mum\ component and 11.2~\mum\ shoulder superimposed on 
the amorphous silicate emission feature at 10~\mum, is very 
unusual when compared to evolved Galactic sources.  Among
the possible explanations available to us, the most plausible
is the presence of crystalline silicates, particularly 
forsterite.  As Figures 5 and 6 show, forsterite can fit the
observed spectral emission well.

Crystalline silicates are well known in the vicinity of young 
stars and in the optically thick shells around OH/IR stars, 
but their presence in optically thin shells around evolved 
stars has yet to be demonstrated convincingly.  \cite{sl03b} 
investigated optically thin dust shells around evolved stars
in the Galaxy, and they found that semi-regular variables,
which have lower mass-loss rates than similar Mira variables,
show stronger emission in the 13~\mum\ feature and the 
correlated dust features at 20 and 28~\mum.  They suggested 
that these dust features may arise from a mixture of 
crystalline alumina (at 13~\mum) and crystalline silicates 
(20 and 28~\mum). 

\cite{gs98} argue that the degree of crystallinity in a 
forming grain is a balance between the timescale for atoms to 
jump into a low-energy location in the lattice structure and 
the rate at which a grain accumulates more atoms.  If the 
grains are forming quickly, then they are accumulating 
material faster than it can settle into the crystalline 
lattice, resulting in an amorphous structure.  However, if
the grains are forming slowly, then each atom has sufficient 
time to find its way into the lattice structure before more 
atoms pile on top of it and lock it into position.  
\cite{kou94} proposed a similar mechanism for the condensation 
of crystalline ice onto grain mantles.  \cite{sl03b} suggested 
that this mechanism could explain the unusual spectral 
structure they observed in semi-regular variables in the 
Galactic sample.

If low mass-loss rates can lead to the direct formation of
crystalline silicate grains, then the generally lower 
metallicities in the Magellanic Clouds should lead to a
higher degree of crystallinity in the outflows from evolved
stars.  Given two stars with the same mass-loss rate but
different metallicities, the star with lower metallicity has
fewer heavy atoms from which dust can form in its outflows,
and as far as dust production is concerned, has a lower
effective mass-loss rate.  The possibility that stars with
lower metallicities might produce crystalline dust is what
makes the spectrum of HV 2310 so fascinating.  

The spectral structure in the 10~\mum\ range shows 
similarities to both the laboratory measurements of 
forsterite and the spectra from galactic spectra which show 
the 13~\mum\ feature (although HV 2310 does not show this 
feature).  The similarity with some galactic spectra 
suggests that at least some of the spectra classified as
structured silicate emission sources (SE4--6)
have enhanced amounts of crystalline silicates 
contributing to the spectral structure in the 10--11~\mum\ 
range.  HV 2310 simply has a {\it higher} degree of
crystallinity than its galactic counterparts.

Because crystalline silicates produce more distinct and 
sharper features than amorphous silicates, their presence
in an observed spectrum makes it easier to analyze the 
properties of the grains.  The spectrum of HV 2310 is 
certainly intriguing, but the relatively high temperature of 
the grains limits our ability to detect crystalline features 
beyond the 10~\mum\ complex, and this in turn limits our
study of the astromineralogy.  The
spectrum does point to the potential for interesting 
discoveries as Spitzer continues to obtain more spectra 
from dust shells around evolved stars in the Magellanic 
Clouds.

\acknowledgements 

We thank Bruce Draine and Dave Harker for useful comments 
on our method of analysis.  We would also like to express
our appreciation to the anonymous referee, whose comments
have greatly improved this manuscript.  The observations 
were made with the Spitzer Space Telescope, which is 
operated by JPL, California Institute of Technology under 
NASA contract 1407 and supported by NASA through JPL 
(contract number 1257184).  This research has made use of 
the SIMBAD and VIZIER databases, operated at the Centre de 
Donn\'{e}es astronomiques de Strasbourg, and the Infrared 
Science Archive at the Infrared Processing and Analysis 
Center, which is operated by JPL.

\end{document}